\documentclass[10pt,conference]{IEEEtran}

\usepackage{amsmath,amsfonts}
\usepackage{graphicx}
\usepackage{booktabs}
\usepackage{caption, colortbl}
\usepackage{multirow}
\usepackage{xspace}
\usepackage{enumitem}
\usepackage{tabularx}
\usepackage{tablefootnote}
\usepackage{tikz}
\usepackage{footnote}
\makesavenoteenv{tabular}
\makesavenoteenv{table}

\usepackage[dvipsnames]{xcolor}

\usepackage{cite}

\usepackage[normalem]{ulem}   
\usepackage{cancel}           

\usepackage[hidelinks]{hyperref}

\usepackage{algorithm}
\usepackage[noend]{algpseudocode}

\usepackage{fancyhdr}
\fancypagestyle{firstpageheader}{
  \fancyhf{}
  
  \fancyhead[C]{\normalsize\bfseries
    Accepted Paper at the IEEE International Symposium on Hardware Oriented Security and Trust (HOST) 2026, Washington, D.C., USA.}
}

\definecolor{RevisionBlue}{RGB}{0,70,200} 

\usepackage[most]{tcolorbox}
\newtcolorbox{blueobs}{
  enhanced, breakable,
  colback=blue!6,
  colframe=blue!60!black,
  boxrule=0.8pt,
  arc=0pt, outer arc=0pt,
  left=6pt, right=6pt, top=6pt, bottom=6pt
}

\newif\ifshowchanges
\showchangestrue 

\ifshowchanges
  \newcommand{\rev}[1]{{\color{black}#1}}
  
  \newcommand{\revdel}[1]{{\color{Gray}\sout{#1}}}
  \newcommand{\revdelmath}[1]{{\color{Gray}\cancel{#1}}}
  \newenvironment{revblock}{\begingroup\color{black}}{\endgroup}
\else
  \newcommand{\rev}[1]{#1}
  
  \newcommand{\revdel}[1]{}
  \newcommand{\revdelmath}[1]{}
  \newenvironment{revblock}{}{}
\fi

\newcommand{\circled}[1]{\tikz[baseline=(char.base)]{
  \node[shape=circle,draw=black,fill=white,text=black,inner sep=0.5pt] (char) {#1};}}

\title{InterPUF: Distributed Authentication via \rev{Physically Unclonnable Functions and Multi-party Computation} for Reconfigurable Interposers}

\author{\IEEEauthorblockN{Ishraq Tashdid\textsuperscript{1}, Tasnuva Farheen\textsuperscript{2}, Sazadur Rahman\textsuperscript{1}}
\IEEEauthorblockA{\textsuperscript{1}University of Central Florida, Orlando, FL, USA, \textsuperscript{2}Louisiana State University, Baton Rouge, LA, USA\\
\{ishraq.tashdid, sazadur.rahman\}@ucf.edu\textsuperscript{1}, tfarheen@lsu.edu\textsuperscript{2}.}}

\begin{document}
\maketitle
\thispagestyle{firstpageheader} 

\begin{abstract}
Modern system-in-Packages (SiP) platforms are rapidly adopting reconfigurable interposers to enable plug-and-play chiplet integration across heterogeneous multi-vendor ecosystem. However, this flexibility introduces severe trust challenges, as traditional security countermeasures fail to scale or adapt in these decentralized, post-fabrication programmable environments. This paper presents \textsc{InterPUF}, a compact, scalable authentication framework that transforms the interposer into a distributed root of trust. At its core, \textsc{InterPUF} embeds a route-based differential delay Physical Unclonable Function (PUF) across the reconfigurable interconnect and secures its evaluation using multi-party computation (MPC). The proposed architecture introduces only $0.23\%$ area and $0.072\%$ power overhead across diverse chiplets while preserving authentication latency within tens of nanoseconds. Simulation results using PyPUF confirm strong uniqueness, reliability, and modeling resistance, even under process, voltage, and temperature variations. By fusing hardware-based PUF primitives with cryptographic hashing and collaborative verification, \textsc{InterPUF} enforces a minimal-trust model without relying on any centralized anchor.


\end{abstract}

\begin{IEEEkeywords}
Secure Heterogeneous Integration, Physically Unclonable Function.
\end{IEEEkeywords}

\vspace{-0.1in}
\section{Introduction}
\vspace{-0.1in}

Cost increases from IC manufacturing processes, shrinking technology nodes, and the growing complexity of integrated circuits have forced original equipment manufacturers (OEMs) to outsource parts of their design, integration, and fabrication to third-party vendors, contract foundries, and packaging facilities. In contrast, a full-custom ASIC allows a single design house to control specification, design, and manufacturing maximizing flexibility and trust~\cite{tehranipoor2011introduction} but at the cost of long development cycles and high non-recurring engineering (NRE) expenses. Foundry-provided process design kits (PDKs) mitigate these costs and improve yield but reduce design flexibility and security. Today’s System-on-Chip (SoC) designs rely heavily on reusable IP blocks, making full custom design impractical. This reuse-driven paradigm has accelerated innovation and reduced both cost and time-to-market. The microelectronics industry is now extending this modular approach through heterogeneous integration, using chiplets, interposers, and advanced 2.5D/3D packaging~\cite{2.5_3D_hetero}. In these system-in-package (SiP) architectures, pre-fabricated logic, memory, analog, and accelerator chiplets from diverse vendors are assembled on shared interposer platforms. Passive and active interposers~\cite{tsv_interposer,cost-effective_passiveinterposer,tashdid2025beyondppa, tashdid2025ecologic, dewan2025FHE, noc_interposer} enable dense, high-bandwidth chiplet communication, while emerging reconfigurable interposers further enable plug-and-play integration~\cite{li2024lucie,li2022gia,Jiao2024FPIA}. With these programmable fabrics, designers can mix and match chiplets like lego blocks, extending product lifetimes through reconfiguration and rapidly prototyping new designs.

\begin{figure}[t]
    \centering    
    \includegraphics[width=\linewidth]{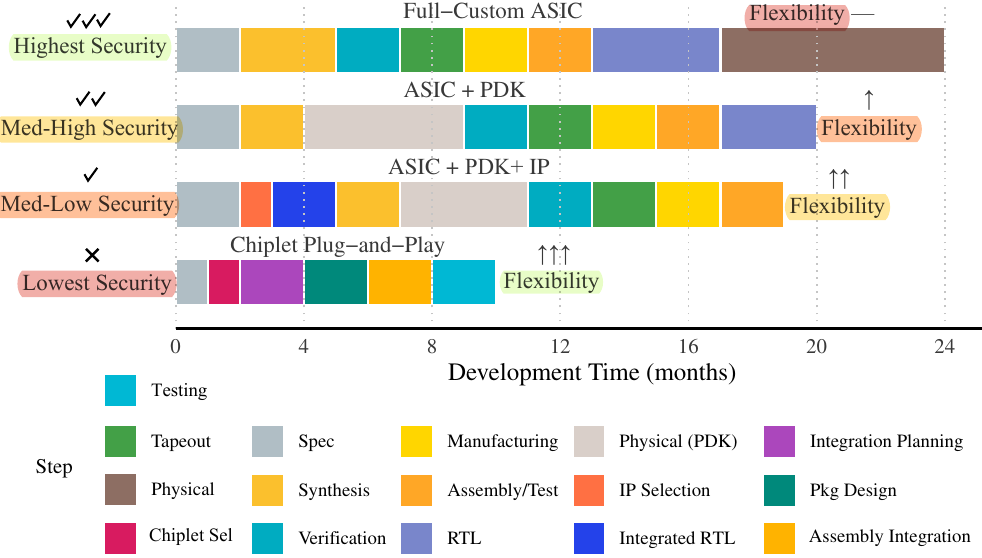}
    \caption{\small Comparison of development time across design flows. Each bar illustrates the sequence and duration of design stages. As it evolves from full-custom to chiplet-based integration, redundant steps e.g., Register Transfer Level (RTL) design or physical implementation are reduced or 
    eliminated, shortening time-to-market.}
    \label{fig:chiplet_ttm}
\end{figure}
The transition to a horizontally distributed IC supply chain, involving multiple global entities, has introduced a range of security threats, including IP piracy, hardware Trojan insertion, counterfeiting, reverse engineering, and overproduction~\cite{shakya2017introduction}. In contrast, earlier full-custom design flows kept specification, design, and fabrication in-house, preserving confidentiality and minimizing exposure to external risks~\cite{rahman2020defense}. As design and manufacturing responsibilities have shifted to outsourced foundries, IP vendors, and chiplet suppliers, the attack surface has expanded dramatically~\cite{toshi}. The threat model for each stage of the design flow is summarized in Table~\ref{tab:micro_tm}. IP vendors may conceal malicious logic~\cite{rahman2020defense}; foundries can overproduce, reverse engineer, or tamper with layouts~\cite{rahman2021ll,postfab_lle}; and integrators must guard against counterfeit or reused chiplets~\cite{farahmandi2023cad}. Furthermore, many SoC-level defenses such as logic locking~\cite{rahman2020defense}, IC metering~\cite{tehranipoor2011introduction}, and scan-chain protections~\cite{rahman2021security} do not directly apply to chiplet-based SiPs, where integrators cannot modify third-party components. \textit{Each outsourcing step thus trades security assurance for faster design cycles, yielding a fragmented, minimal-trust supply chain.}


Reconfigurable interposer–based SiPs most starkly expose the speed–trust trade-off: programmability enables adaptive routing and interoperability yet centralizes attack surface~\cite{biswas2025sample}. Thus, the interposer must serve as a verifiable root of trust or modularity becomes systemic risk. \begin{revblock}
Existing SiP trust schemes rely on centralized anchors (a trusted chiplet/port)~\cite{gate_sip,sect_hi}, creating bottlenecks and single points of failure; cryptographic handshakes add latency~\cite{pqc_hi}; many MPC approaches still assume a central validator~\cite{tashdid2025safe}; and interposer-centric RoTs remain fixed and don’t scale across reconfigurable, multi-vendor meshes~\cite{Nabeel2019InterposerRoT}. We therefore need distributed, scalable trust. MPC enables joint verification without revealing secrets~\cite{tashdid2025safe, tashdid2025authentree, mpcircuits,tinygarble} and is proven at scale in software~\cite{software_mpc,mpc_ml,Halevi2021}. We propose \textsc{InterPUF}: a decentralized, interposer-resident trust fabric that verifies authenticity via MPC while never exporting raw PUF outputs. Our main contributions follow.
\end{revblock}
\begin{itemize}[leftmargin=*]
\item We propose \textsc{InterPUF}, which embeds a differential, route-based delay PUF in the reconfigurable interposer and uses it as the SiP-wide root of trust.

\item We design a protocol that secret-shares only stabilized tokens and chiplet encryption, enabling distributed accept/reject decisions without a central validator with raw challenge–response pairs (CRPs) never cross the boundary.

\item Our hardware evaluation shows that \textsc{InterPUF} achieves negligible interposer overhead below $0.23\%$, per–chiplet encryption overhead as low as $0.48\%$, and power under $0.072\%$, while maintaining fast authentication latency.

\item Simulation results with \textsc{pyPUF} confirm strong PUF quality (uniformity $\sim0.5$, reliability $98\%$, uniqueness $\sim0.46$), and machine-learning attacks achieve only random-guess accuracy ($\sim47\%$), validating resilience.

\item We analyze resilience to modeling, replay, probe, and \begin{revblock}Denial of Service (DoS)\end{revblock} attacks under a sceptical-trust supply chain.

\item To accelerate further research and facilitate industrial adoption, we will release our Verilog implementation, evaluation scripts, and benchmark data as open-source at our GitHub.
\end{itemize}

\label{sec:intro}

\noindent The paper is organized as follows: Sec.~\ref{sec:background} reviews related work, Sec.~\ref{sec:threat} defines the threat model, and Sec.~\ref{sec:overview} describes the architecture and operation of \textsc{InterPUF}. Sec.~\ref{sec:modeling} outlines the simulation flow, Sec.~\ref{sec:eval} reports evaluation results, and Sec.~\ref{sec:security} analyzes resilience. Sec.~\ref{sec:discussion} discusses limitations, and Sec.~\ref{sec:conclusion} concludes.

\begin{table}[t]
\vspace{-3pt}
\centering
\caption{Threat model across common design flows.\tablefootnote{\begin{revblock} Here `+' means addition along with the previous entries. For example, ASIC+PDK+IP design flow involves design house, foundry, and vendors while threat model includes insiders, untrusted foundry, and IP piracy. Common threats, such as, hardware Trojans; side-channel leakage, fault injection, test/DFT abuse, forged documentation, are omitted for space.\end{revblock}}}
\label{tab:micro_tm}
\setlength{\tabcolsep}{9pt}
\resizebox{0.5\textwidth}{!}{%
\begin{tabular}{lll}
\toprule
\textbf{Design Flow} & \textbf{Entities Involved} & \textbf{Threat Models} \\
\midrule
Full-custom ASIC        & Design house                          & Insider threat \\
ASIC+PDK                & + Foundry                              & + Untrusted Foundry \\
ASIC+PDK+IP             & + Vendors                               & + IP Piracy \\
Chiplet Plug-and-Play   & + Packaging                             & + Chiplet Foundry \\
\bottomrule
\end{tabular}
}
\vspace{-6pt}
\end{table}

\section{Background and Related Work}\label{sec:background}
Authentication in heterogeneous multi-chiplet systems requires both architectural support and scalable trust. This section reviews prior work along three directions: \circled{\small 1} advances in heterogeneous integration and reconfigurable interposers enabling plug-and-play assembly; \circled{2} existing techniques for SiP security, which often rely on centralized anchors or heavy cryptography; and \circled{\small 3} delay-based PUFs, a compact primitives for unique hardware signatures, motivating \textsc{InterPUF}.
\vspace{-9pt}
\subsection{Heterogeneous Integration and Reconfigurable Interposers}\label{subsec:hi_and_reconfig}
Heterogeneous Integration (HI) is the cornerstone of microelectronics in the post-Moore era by treating the interposer not as a passive carrier but as an active design resource that enables modularity at scale~\cite{lau2021chiplet}. This modular approach reduces cost, improves yield, and shortens design cycles by reusing known-good dies across multiple products. However, today’s interposers are mostly custom-designed for each system, limiting reuse and constraining agility~\cite{li2022gia}. \begin{revblock}In practice, most SiP interposers still employ fixed metal routing; truly reconfigurable routing meshes remain an active research direction rather than commodity practice~\cite{li2024lucie,Ehrett2019Sipterposer,Jiao2024FPIA,li2022gia,thermal_reusable}. By decoupling chiplet function from wiring, these techniques enable plug-and-play composition (late binding, binning, and field repair), but they also introduce new security challenges as highlighted next.\end{revblock} 
\vspace{-9pt}
\subsection{Challenges of Reconfigurable Interposers}\label{subsec:reconfig_challenges}

\rev{Reconfigurable silicon interposers, while offering in-field programmability, also enlarge the attack surface.}~Unlike static interposers, their dynamic reprogrammability can introduce new runtime vectors for adversaries to exploit post-deployment~\cite{johnson2017remote, chakraborty2013hardware}. \circled{\small 1} Adversaries may reconfigure the interposer to inject hardware Trojans or alter interconnect topologies to subvert system behavior~\cite{charles2021survey,chakraborty2013hardware}. \circled{\small 2} Moreover, unauthorized access to communication channels may hijack or reroute data flows across chiplets~\cite{charles2021survey}. And, \circled{\small 3} covert side-channel leakage, enabled by dynamically reconfiguring signal paths to observe or modulate sensitive data~\cite{bommana2025mitigating}. 
\rev{Hence, ensuring provenance across evolving configurations, therefore, requires attestation of configuration state, secure firmware for the configuration controller, and lifecycle monitoring.}
\vspace{-9pt}


\begin{table}[t]
\caption{Contrast with Prior Authentication Approaches.}
\centering
\label{tab:intro_comparison_alt}
\resizebox{0.48\textwidth}{!}{
\setlength{\tabcolsep}{0.1pt}
\begin{tabular}{ccc}
\toprule
\textbf{Method} & \textbf{Drawbacks} & \textbf{Advantages of \textsc{InterPUF}} \\
\midrule
\textbf{GATE-SiP~\cite{gate_sip}} &
\begin{tabular}[c]{@{}c@{}}TAP-based;\\susceptible to MITM\end{tabular} &
\begin{tabular}[c]{@{}c@{}}No TAP changes;\\robust against MITM\end{tabular} \\
\midrule
\textbf{PQC-HI~\cite{pqc_hi}} &
\begin{tabular}[c]{@{}c@{}}High computation cost;\\leakage under probing\end{tabular} &
\begin{tabular}[c]{@{}c@{}}low-overhead, distributed\\checks with strong\\signature hashing\end{tabular} \\
\midrule
\textbf{SECT-HI~\cite{sect_hi}} &
\begin{tabular}[c]{@{}c@{}}Focuses test encryption;\\limits vendor flexibility\end{tabular} &
\begin{tabular}[c]{@{}c@{}}Supports both vendor\\and integrator security\end{tabular} \\
\midrule
\textbf{SAFE-SiP~\cite{tashdid2025safe}} &
\begin{tabular}[c]{@{}c@{}}Relies on a central hub;\\ \end{tabular} &
\begin{tabular}[c]{@{}c@{}}Distributed validation;\\no single point of failure;\\scalable and efficient\end{tabular} \\
\bottomrule
\end{tabular}}
\vspace{1mm}

\raggedright \small
TAP: Test Access Port. \quad
MITM: Man-in-the-Middle. \\

\end{table}

\subsection{Existing Works and Their Drawbacks}\label{subsec:existing_techniques}
\rev{Prior works (Tab.~\ref{tab:intro_comparison_alt}) focus on static interposers via fabrication-time countermeasures, hardware primitives, and cryptographic protocols. Split manufacturing (SM) and network-on-interconnect (NoI) obfuscation conceal design information to impede reverse engineering~\cite{splitcore}. SM’s reliance on trusted back-end fabrication limits scalability and compromises yield, while NoI obfuscation complicates integration and validation~\cite{postfab_lle}. Module-level approaches, e.g., Chiplet Hardware Security Modules (CHSM) and Chiplet Security IP (CSIP), provide authentication but at the cost of additional area and design complexity, and they (re)introduce single points of failure~\cite{toshi}. Researchers proposed post-quantum cryptographic authentication and encrypted test protocols~\cite{pqc_hi,sect_hi} that rely on centralized trust anchors and incur non-trivial latency overheads. Nonetheless, MPC-assisted frameworks~\cite{tashdid2025safe} are typically coordinator-based. Moreover, none of these methods explicitly target reconfigurable interposers, where routing state can evolve over time (shown in Sec.~\ref{subsec:hi_and_reconfig},~\ref{subsec:reconfig_challenges}). Collectively, these limitations motivate an authentication primitive that is efficient, scalable, and trust-minimized/coordinator-free, and that is tailored to reconfigurable interposer fabrics.}
\vspace{-9pt}
\subsection{Delay-based PUFs}\label{subsec:delay_puf}
Physical Unclonable Functions (PUFs) include several families:
\circled{a} Arbiter PUFs, where challenge bits steer two matched delay paths and an arbiter records the race—yielding many challenge–response pairs (CRPs) but making them vulnerable to modeling and side-channel attacks when CRPs are exposed~\cite{lee2004technique,lim2005extracting};
\circled{b} Ring-oscillator (RO) PUFs, which compare oscillator frequencies and are well suited for FPGAs~\cite{suh2007physical};
\circled{c} Between-die delay PUFs, extending delay races across chips through bumps, vias, and package traces~\cite{rosenfeld2010sensor}; and
\circled{d} Interconnect PUFs, which embed races in interposer or routing fabrics for SiP authentication~\cite{yu2019interconnect}.
\begin{revblock}
Arbiter PUFs typically chain $64$–$512$ stages to build entropy. Delay-based PUFs are sensitive to process, voltage, temperature, and aging variations, which can be mitigated through majority voting, bit pre-selection, and lightweight error correction. Modeling attacks grow effective with large CRP exposure, while side channels (power, timing, EM) can leak responses; countermeasures include limiting CRPs, hashing challenges, and retaining raw responses on-chip.
\end{revblock}

\begin{figure}[t]
    \centering
    \includegraphics[width=\linewidth]{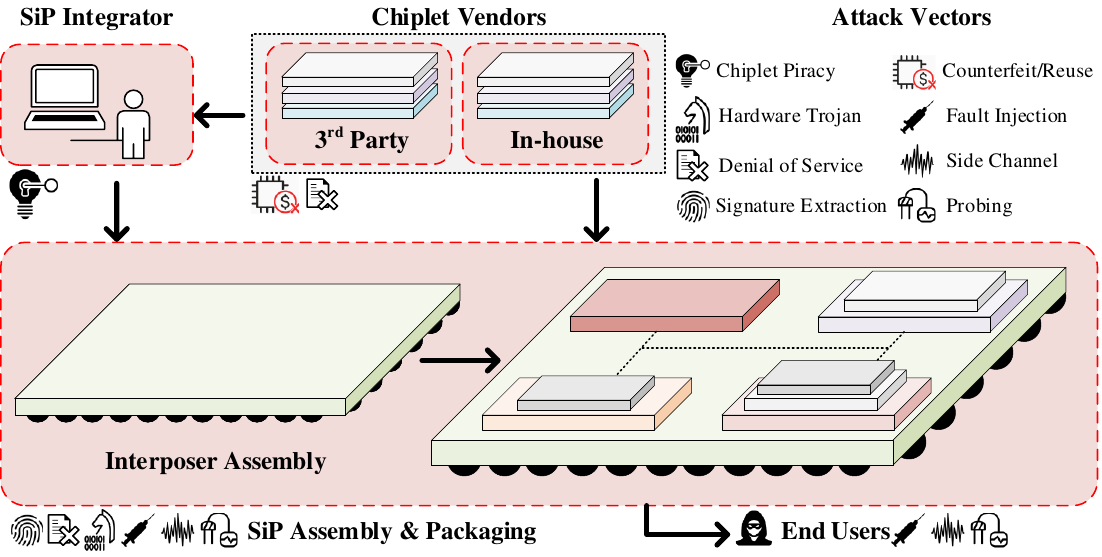}
    \caption{\small SiP design flow and life-cycle. Untrusted and trusted parties are marked by red and green boxes, respectively.}
    \label{fig:threatmodel}
\end{figure}

\section{Threat Model and Assumptions}\label{sec:threat}


Our framework adopts a \rev{minimal-trust} assumption in which no single actor in the heterogeneous integration flow is \rev{fully} trusted~\cite{stern2021aced}. The relevant actors are \circled{a} chiplet vendors (counterfeit or reused chiplets), \circled{b} the interposer/packaging foundry (tampering with routing fabrics, unauthorized reconfiguration, interconnect probing), \circled{c} system integrators (semi-honest or malicious misuse of authentication artifacts), and \circled{d} end-users/field adversaries (runtime reconfiguration after deployment). Chiplets are treated as identity-bearing components with opaque internals; we do not verify internal functionality. \rev{Moreover, out-of-scope threats include malicious-by-design chiplets with functional Hardware Trojans; such cases require additional supply-chain defenses beyond our focus.}

\rev{Parties participating in verification are at least semi-honest (follow the protocol but may be curious). \rev{Enrollment} occurs in a trusted environment or procedure to obtain reference data (e.g., golden challenge–response information). \rev{Raw PUF responses are not exported during operation}, and reference use avoids exposing raw CRPs. \rev{Attack capabilities} considered include \circled{i} cloning/reuse and substitution, \circled{ii} replay and chosen-challenge attempts aimed at modeling the PUF, \circled{iii} probing or rerouting on the interposer to extract or bias signals, and \circled{iv} configuration tampering and \rev{denial-of-service (DoS)} by abusing reprogrammability. \rev{Side-channel attacks} (power, EM, timing) are acknowledged but full side-channel hardening is beyond the scope of this work and discussed as a limitation.}

\noindent Given the adversary model, our security objectives include- 
\begin{itemize}[leftmargin=*]
\item Authenticate chiplets without exposing raw CRPs. 
\item Resist modeling, replay, and cloning attacks even under partial compromise of the supply chain. 
\item Maintain robustness to moderate noise, environmental variations, and aging effects. 
\item Scale efficiently with mesh size and chiplet count in large SiP assemblies.
\end{itemize}

\begin{figure}[t]
\centering
\includegraphics[width=\linewidth]{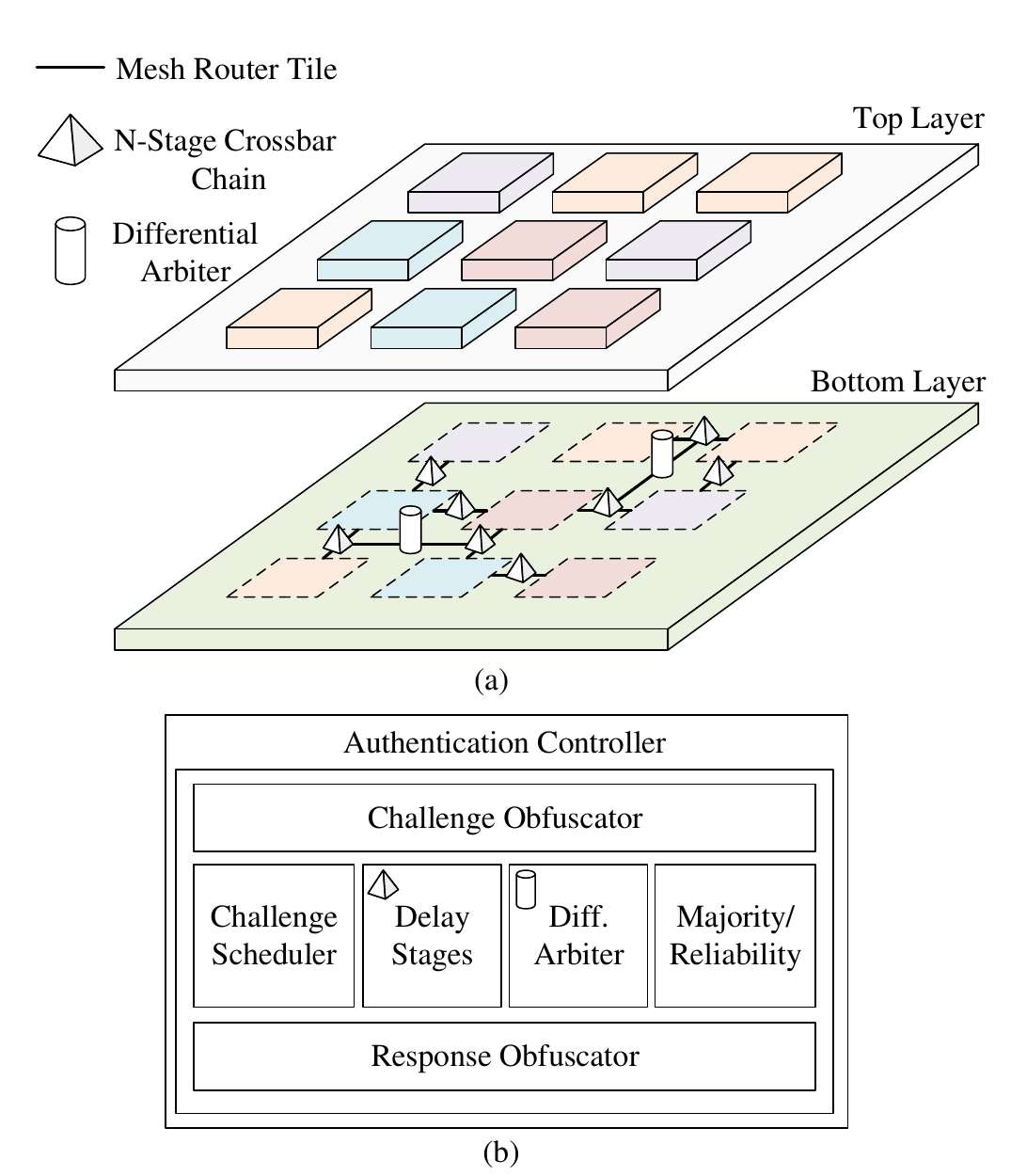}
\caption{Architecture of \textsc{InterPUF}. (a) Physical organization of the system, where the bottom layer integrates the interconnect mesh with embedded $N$-stage crossbar chains and differential arbiters, and the top layer hosts chiplets interconnected through the fabric. (b) Authentication controller, which manages challenge hashing, scheduling, differential arbitration, majority/reliability voting, and response hashing.}
\label{fig:interpuf_architecture}
\end{figure}

\section{System Overview}
\label{sec:overview}
In this section, we present the \textsc{InterPUF} architecture which leverages interposer-integrated routing, compact delay chains, and distributed chiplet wrappers to enable scalable authentication in multi-chiplet systems. Sec.~\ref{subsec:arch} elaborates on the high-level system overview, illustrated in Fig.~\ref{fig:interpuf_architecture} while ~\ref{subsec:operation} discusses the system-level operation, shown in Fig.~\ref{fig:interpuf_flow}.

\subsection{Architecture} \label{subsec:arch}

This subsection provides a high-level overview of our architecture which will be made publicly available on GitHub after publication. While a PUF is inherently a physical phenomenon that cannot be fully captured through logical code, we approximate its behavior here and later present a detailed simulation-based model in Sec.~\ref{sec:modeling}.

\noindent\textsc{\textbf{\underline{1. Interposer Mesh:}}}
At the heart of the system lies a Manhattan mesh interconnect, the plug-and-play base proposed in~\cite{li2024lucie} for 2.5D interposer-based integration. The mesh is composed of router tiles, each supporting bidirectional links to adjacent tiles. This topology is both cost-effective and reconfigurable, minimizing wiring overhead while maintaining sufficient flexibility for large-scale integration. Every router tile includes a compact switchbox that determines the direction of signal flow. The simplicity of this design not only reduces area and power but also allows low-overhead PUF integration at router boundaries. \uline{As a result, the interconnect itself becomes an active participant in system-level authentication rather than a passive wiring fabric.}

\begin{figure*}[t]
    \centering
    \includegraphics[width=\linewidth]{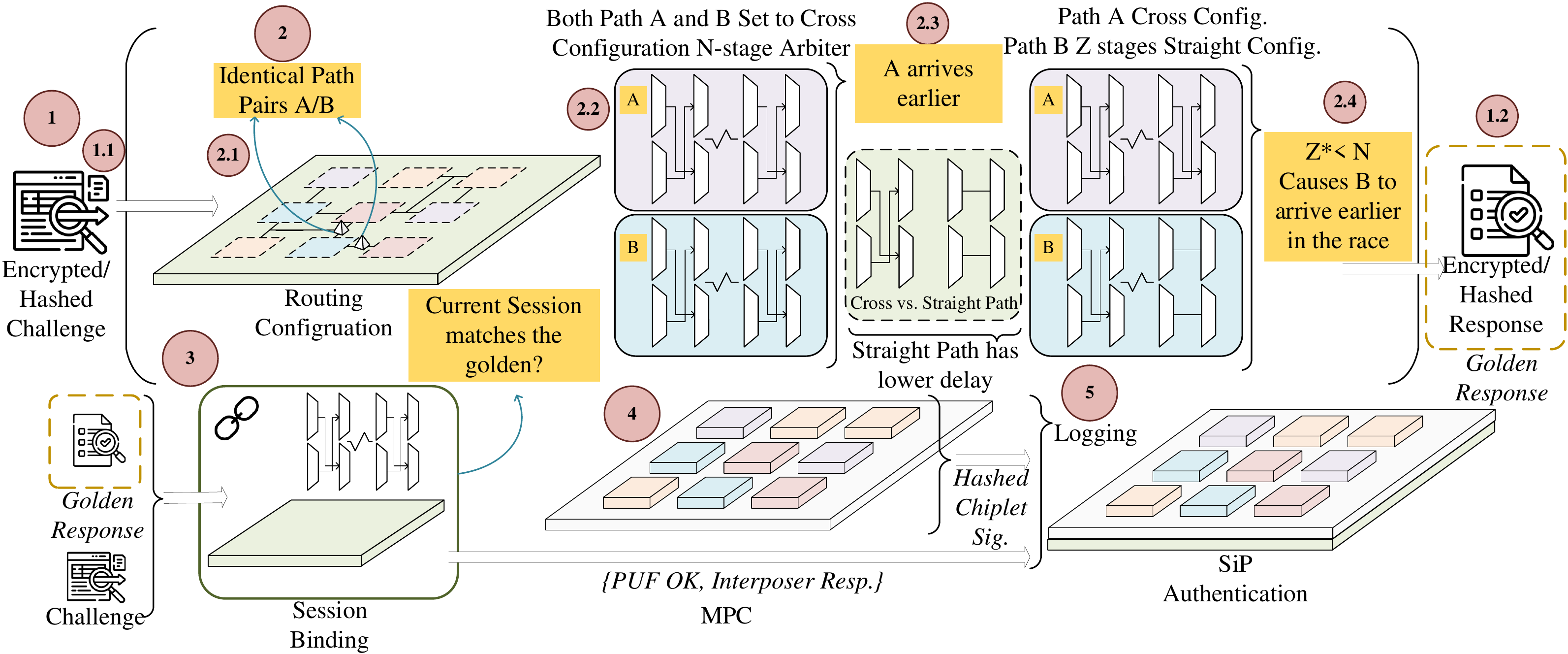}
    \caption{\rev{Operation flow of \textsc{InterPUF}.}}
    \label{fig:interpuf_flow}
\end{figure*}

\noindent\textsc{\textbf{\underline{1.a Process:}}} In \textsc{InterPUF}, each router tile embeds a small ``$N$-Stage'' crossbar-based delay chain, which is the base of the PUF. A pair of neighbor routes is selected per challenge, forming an ``A-path'' and a ``B-path'' across the mesh. The delays along these two paths are compared using a differential arbiter located at the sink. Sec.~\ref{subsec:operation}-2 shows an example of this path pair. Since process variation ensures unique delay characteristics per device, the arbiter’s outcome serves as a device-specific response. Repeating this process across challenges yields a rich response space that is tightly bound to the physical structure of the interconnect. 
\rev{Therefore, we instantiate many neighbour or identical length source--sink pairs and, per pair, each stable race contributes one bit, so aggregating across pairs and permutations yields a $256$- or $512$-bit route digest without requiring $256/512$ dedicated stages in a single path. Only challenges that pass stability thresholds across PVT. All PUF stages (delay elements and switchboxes) reside on the interposer, chiplets act only as sources/sinks, and the arbiter is time-multiplexed at the interposer edge. For scale, a $16{\times}16$ mesh with typical source--sink distances of $20$--$30$ hops and $40$--$80$ path pairs, each with $8$ path permutations, produces $320$--$640$ candidate bits; after stability filtering ($\sim$60--80\%), a $256$ bit digest is readily realized without on-chip stages.}

\begin{revblock}
    
\noindent\textsc{\textbf{\underline{1.b Golden-Free Interposer Self-Check}}}
To detect route tampering without any external golden reference, the interposer exploits the fact that a crossed switchbox setting introduces more delay than a straight setting~\cite{puf_straight}. Consider an $N$-stage differential path race between two routes (A and B) connecting neighboring chiplets (See Fig.~\ref{fig:interpuf_flow}-2.1). We begin from a symmetric baseline where both routes are configured all-cross (Fig.~\ref{fig:interpuf_flow}-2.2). Suppose A arrives earlier (Fig.~\ref{fig:interpuf_flow}-2.3), indicating the winner of the race. Then we change the configuration of route~B by flipping the first $Z$ stages from cross to straight (keeping all stages on route~A as cross) and evaluate the arbiter outcome again under majority voting (Fig.~\ref{fig:interpuf_flow}-2.4). The smallest $Z$ at which B consistently wins is recorded as the threshold $Z^\star$ for that \{source, sink\} pair.

Intuitively, $Z^\star$ captures how much delay “headroom” exists between the two nominally identical routes. Honest, untampered fabrics produce tightly clustered $Z^\star$ values across different pairs. So, any hidden delay (e.g., a stealth insertion, unintended extra buffering, or re-route) along a route shifts its threshold, making its $Z^\star$, a clear outlier from the population mean $Z_{\text{avg}}$. During enrollment, the controller collects the vector of thresholds $Z^\star$ across many local pairs and stores in the designers secure repository (detail discussed in Sec.~\ref{subsec:operation}.2). In the field, periodic re-checks recompute $Z^\star$ and flag routes whose thresholds drift beyond the allowed band, providing \uline{golden-free} integrity monitoring. If a dormant Trojan was inactive at enrollment~\cite{2023trojan_detection}, later periodic checks (performed opportunistically at boot or during low-load windows) still surface its effect when it activates, as the corresponding $Z^\star$ deviates sharply from $Z_{\text{avg}}$.

\end{revblock}

\noindent\textsc{\textbf{\underline{2. Authentication Controller:}}} \textit{In \textsc{InterPUF}, authentication is not pushed down to the chiplets via dedicated wrappers, ensuring that even if third-party vendors performs denial-of-service, the scheme cannot be bypassed since authentication remains a centralized task enforced at the interposer level.} To ensure this, the authentication controller embedded in the interconnect is composed of several low-overhead but coordinated components as shown in Fig.~\ref{fig:interpuf_architecture}.
\begin{itemize}
    \item A \emph{challenge scheduler}, which orchestrates and delivers challenges across the mesh fabric.
    \item A \emph{challenge obfuscator}, which transforms challenges to conceal any correlations, increasing modeling resistance.
    \item Distributed delay stages and differential arbitrators, which evaluate timing races induced by challenges.
    \item A \emph{majority/reliability unit}, which repeats evaluations, suppresses noise, and filters unstable challenges. This improves validity against PVT variation, and also against adversarial perturbations, such as hardware Trojans that may selectively trigger faults in a single instance~\cite{puf_ht}.
    \item A \emph{response obfuscator}, which hashes or encodes final response bits into short authentication tokens before exposing them externally, ensuring that raw delay information is never directly observable.
\end{itemize}

This interconnect-level controller establishes a scalable two-layer trust hierarchy. The interposer fabric itself becomes the first line of defense, validating routing integrity and guaranteeing a trustworthy communication backbone. On top of this, individual chiplets contribute their cryptographic signatures, which are evaluated only after the interconnect has been secured. By clearly decoupling the compact interconnect authentication from heavier chiplet-level encryption (discussed in Section~\ref{subsec:operation}), the system achieves both efficiency and robustness without duplicating logic across every chiplet.
\vspace{-9pt}
\begin{revblock}
    
\subsection{Operation}
\label{subsec:operation}

We present an interposer-centric flow in Fig.~\ref{fig:interpuf_flow} in which (i) the interposer exposes only a golden route digest derived from its differential-delay PUF and (ii) each chiplet binds its identity to that digest at enrollment. At runtime, authentication is realized as a constant-round two-party MPC between the interposer (verifier) and a chiplet (prover).

\noindent\textsc{\textbf{\underline{0. Notation:}}}
Interposer $\mathcal{I}$; chiplets $\mathcal{C}=\{C_1,\ldots,C_N\}$. Let $\mathbf{ch}$ be a routing challenge, $\text{\textsc{Epoch}}$ a session identifier, and $\text{\textsc{Nonce}}$ a fresh per-session value. $\text{\textsc{SHA256}}$ denotes SHA-256; $\text{\textsc{HKDF}}$ is HMAC-based KDF (Key Derivation Function). Concatenation is $\parallel$. $\text{\textsc{PRK}}$ is Pseudo-Random Key.

\noindent\textsc{\textbf{\underline{1. Challenge-Response:}}}
During enrollment, for a curated challenge set and PVT corners, $\mathcal{I}$ evaluates differential paths with $K$-fold repetition and majority vote (Fig.~\ref{fig:interpuf_flow}-1.1).  Responses with flip-rate below $\tau$ are retained. The accepted bitstring is hashed to a fixed-length golden route digest, $R^\star$, with the help of a designer chiplet (Fig.~\ref{fig:interpuf_flow}-1.2).
\vspace{-6pt}
\[
  R^\star \;=\; \text{\textsc{SHA256}}(\text{stable\_PUF\_bits}).
  \vspace{-6pt}
\]
Raw CRPs never leave $\mathcal{I}$; only $R^\star$ is exported for binding.

\noindent\textsc{\textbf{\underline{2. Enrollment:}}}
Enrollment runs on standard Automated Test Equipment (\textsc{ate}) at the Outsourced Semiconductor Assembly and Test (\textsc{osat}). The \textsc{ate} executes a signed test program, and the tester host is paired with a Hardware Security Module (\textsc{hsm}), possibly a random designer chiplet, that attests tester identity, gates cryptographic operations, and produces signed logs. But both parties need to exchange hash to confirm the security of the process. For a curated challenge set and representative PVT corners, the interposer evaluates many races, and hashes the admitted bits to produce the golden route digest $R^\star$. Optional helper data (if used) are stored as non-sensitive metadata in on-device Non-Volatile Memory (\textsc{nvm}). Each digest is bound to traceability fields (device identifier, wafer, lot), and the \textsc{hsm} signs a per-device enrollment manifest (device ID, digest set/version, trace metadata, tester attestation) that is uploaded to the designer’s secure repository. The \textsc{osat} retains only pass/fail and non-sensitive logs. Separately, each chiplet $C_i$ binds its local identity to the interposer’s \textsc{puf} via a one-time commitment
  \vspace{-12pt}

\[
G_i=\text{\textsc{sha256}}\!\big(\text{\textsc{ID}}_i \,\|\, \text{\textsc{SIG}}_i \,\|\, R^\star \,\|\, \text{\textsc{EnrollTag}}\big),
  \vspace{-6pt}
\]
where $\text{\textsc{ID}}_i$ is a device-local identifier and $\text{\textsc{SIG}}_i$ is a vendor signature or silicon-resident secret attesting $\text{\textsc{ID}}_i$. Only $G_i$ and trace metadata are provided to the integrator; $(\text{\textsc{ID}}_i,\text{\textsc{SIG}}_i)$ never leave $C_i$. The interposer stores $G_i$ (and the signed manifest) in secure \textsc{nvm}.

\noindent\textsc{\textbf{\underline{3. Session Binding:}}}
The interposer derives a per-session salt \(s\) from the golden digest \(R^\star\) and the current context \((\mathbf{ch},\text{\textsc{Epoch}})\), and samples a fresh \text{\textsc{Nonce}}; both values are public within the session and bind all proofs to this interposer instance while preventing replay. Authentication proceeds only if the current challenge passes the PUF stability check (\text{\textsc{puf\_ok}}{=}1) (Illustrated in Fig.~\ref{fig:interpuf_flow}-3).

\noindent\textsc{\textbf{\underline{4. MPC:}}} The goal is to decide if the chiplet holds identity material that reproduces $G_i$ bound to $R^\star$, and produce a per-session token tied to $(s,\text{\textsc{Nonce}})$, without revealing either party’s secrets. Fig.~\ref{fig:interpuf_flow}-3 shows that in each run, the interposer supplies its private inputs $(G_i, R^\star)$, the chiplet supplies its private inputs $(\text{\textsc{ID}}_i,\text{\textsc{SIG}}_i)$, and both parties use the public inputs $(s,\text{\textsc{Nonce}},\text{\textsc{PUF\_OK}})$. They execute two-party computation using Yao garbled circuits with Oblivious Transfer to evaluate a fixed circuit $f_i$, inspired by~\cite {tashdid2025safe}. 
\begin{itemize}[leftmargin=*]
\item The circuit first recomputes $G'_i=\text{\textsc{sha256}}(\text{\textsc{ID}}_i\parallel\text{\textsc{SIG}}_i\parallel R^\star\parallel\text{\textsc{EnrollTag}})$. 
\item It then sets $b_i$ to true \textit{iff} $G'_i=G_i$ and $\text{\textsc{PUF\_OK}}=1$.
\item It finally derives a session token $T'_i=\text{\textsc{sha256}}(G'_i\parallel s\parallel\text{\textsc{Nonce}})$. Only $(b_i,T'_i)$ are revealed to the interposer. 
\item Concretely, the interposer garbles $f_i$ and sends the garbled tables to the chiplet, provides input labels for $(G_i,R^\star)$, and the chiplet obtains labels for $(\text{\textsc{ID}}_i,\text{\textsc{SIG}}_i)$ via \textsc{ot}. 
\item Both parties encode $(s,\text{\textsc{Nonce}},\text{\textsc{PUF\_OK}})$ as public inputs. 
\item The chiplet evaluates the garbled circuit and returns the garbled outputs, and the interposer decodes them to obtain $(b_i,T'_i)$ and accepts only if $b_i=1$. 
\end{itemize}

\noindent This \textsc{2pc} protocol completes in a constant number of rounds, independent of circuit depth.

\noindent\textsc{\textbf{\underline{5. Logging:}}}
Upon acceptance ($b_i{=}1$), the interposer may record $(i,\mathbf{ch},\text{\textsc{Epoch}},\text{\textsc{Nonce}},T'_i)$ for audit and replay detection, while avoiding any storage of secret inputs. Because $T'_i=\text{\textsc{sha256}}(G_i\!\parallel\!s\!\parallel\!\text{\textsc{Nonce}})$ and the salt $s$ is derived fresh each session, tokens are unlinkable across sessions.

\begin{blueobs}
An impostor that lacks $(\text{\textsc{ID}}_i,\text{\textsc{SIG}}_i)$ bound to $G_i$ and $R^\star$ cannot satisfy $G'_i{=}G_i$ inside the \textsc{2pc} check and therefore cannot produce a valid $T'_i$. Token replay is ineffective because $(s,\text{\textsc{Nonce}})$ changes every session. Copying $G_i$ alone is insufficient without the corresponding identity material. Throughout the protocol, neither $G_i$ nor $(\text{\textsc{ID}}_i,\text{\textsc{SIG}}_i)$ is revealed. For any subset $\mathcal{S}\!\subseteq\!\{1,\ldots,N\}$, the interposer runs $f_i$ independently for each $i\!\in\!\mathcal{S}$ (in parallel when available) and then applies a local policy predicate to the resulting bits $\{b_i\}_{i\in\mathcal{S}}$ (e.g., quorum) to produce the final decision.
\end{blueobs}

\end{revblock}

\section{Modeling and Simulation Methodology}
\label{sec:modeling}

This section presents a complete, interposer-aware simulation flow designed to mirror the register-transfer–level (RTL) implementation. We employ \textsc{pyPUF}~\cite{pypuf} to reproduce parametric PUF behavior, integrate standard machine-learning frameworks for modeling attacks, and introduce pre- and post-processing layers that truly emulate the interconnect transformations and hashing wrappers observed in our RTL.

\subsection{Design–Faithful System Abstraction}
The hardware architecture is based on a Manhattan-style mesh, where each router and interconnect contributes to the PUF’s effective stage depth. In simulation, this is abstracted as the sum of tile sites and horizontal/vertical links, matching the RTL logic. Our default grid size of four yields a stage count consistent with the synthesizable mesh-based design. Each simulated chip instantiates an XOR-of-$K$ arbiter family (with $K\geq 4$) to reflect the nonlinear aggregation introduced in RTL by multiple chains and parallel arbiters. Where the RTL uses tapped feedback to introduce feed-forward complexity, the simulator mimics this with lifted challenge features (pairwise interactions between chosen tap indices). This ensures that the separation and nonlinearity effects captured in silicon are faithfully approximated in the software model.

The interconnect wrapper is modeled in three stages, aligned with how the RTL routes and perturbs challenge bits: (i) permutation of stage indices to emulate reordering by routers, (ii) sparse polarity flips to reflect wiring inversions and mode switches, and (iii) a sparse, invertible binary mixing matrix to capture weak coupling among neighboring links. Each router configuration deterministically defines these transformations, ensuring that every interconnect setting produces a consistent, device-specific challenge mapping. The mapping remains invertible by construction, ensuring that entropy is not artificially added or lost, thereby mirroring the RTL’s structure-preserving routing logic.

\subsection{Measurement Pipeline and PVT Realism}
CRPs are generated in large batches across multiple simulated devices to represent chip-to-chip variability. In hardware, inter-chip differences arise from uncontrollable process variation; in simulation, each chip is seeded independently to induce distinct delay biases. Likewise, interconnect diversity is introduced by random router configurations per challenge batch. To reflect operating conditions realistically, we incorporate Process, Voltage, and Temperature (PVT) effects as low-probability response flips. This models metastability, supply noise, and temperature-dependent delay shifts. Each challenge is evaluated multiple times, and a majority combiner, implemented with the same repetition depth as the RTL, produces a stable bit. Challenges are scored by their stability, and only the most reliable subset is retained, which mirrors enrollment in actual deployments where marginal CRPs near the decision boundary are discarded, ensuring that the dataset passed to the adversary matches what would realistically be exposed by deployed hardware.

\subsection{Controlled Interfaces and Digest}
In line with our RTL, challenges are not applied directly to the delay chain but first passed through an invertible linear transformation over $\mathrm{GF}(2)$. This hashing layer preserves entropy while removing fixed alignment between the external challenge and the internal stage basis, preventing straightforward modeling attacks. In simulation, we apply the same principle: a random but invertible binary matrix transforms challenges before evaluation. 

Raw PUF bits never leave the system unprotected. In RTL, a low-overhead hashing stage converts stabilized responses into authentication tokens; in simulation, we use a 32-bit diffusion-oriented combiner that outputs ${\pm1}$ symbols. Session-bound digest is applied only after reliability filtering, ensuring instability is never masked. As a result, external observers or attackers only see obfuscated tokens rather than raw race outcomes, faithfully mirroring the deployed hardware’s information flow.

\subsection{Attacker Models and Training Protocols}
To evaluate resilience, we consider both oracle and deployed modes. In the oracle mode, all protections are disabled, exposing raw arbiter responses. This provides a conservative lower bound, allowing standard logistic regression on $\phi$-mapped features to model plain arbiter structures. In more realistic configurations (parallel XOR chains, feature-lifted challenges, interconnect mixing), we train multilayer perceptrons with tanh activations sized to the challenge dimension. Training and testing datasets are fixed across experiments to maintain reproducibility. Also, when response tokenization is enabled, direct CRP modeling is not possible, by interface design. In these cases, reported attack accuracy is suppressed not due to modeling limitations but because the system interface simply withholds raw data. This mirrors deployment reality, where external entities never access the unprotected bitstream. Thus, all reported attack results under oracle mode represent strict worst-case bounds, while the deployed mode faithfully mirrors the RTL’s hardened interface.

\subsection{Metric Computation}
We evaluate three canonical PUF metrics: uniqueness (average Hamming distance between devices), uniformity (average fraction of ones per device), and reliability (agreement across repeated reads). Importantly, these are measured \emph{before} tokenization, aligning with how RTL majority voters stabilize responses. When tokenized outputs are used, proxy measurements (bitwise disagreements across tokens) are adopted to maintain comparability, all reported in Sec.~\ref{subsec:puf_eval}.

\section{Evaluation} \label{sec:eval}

This section evaluates the practicality of \textsc{InterPUF} by analyzing its area, timing, and power overhead and compared with other relevant solutions along with results for simulation presented in the previous section. 

\subsection{Experimental Setup}

We implemented \textsc{InterPUF} in Verilog and synthesized the design using Synopsys Design Compiler with the SAED $14\,\mathrm{nm}$ standard cell library. Post-synthesis netlists were used to extract power, area, and timing metrics. All experiments were conducted on a dual-socket Intel Xeon (Skylake) server equipped with $32$ cores and $190\,\mathrm{GB}$ of RAM. The synthesized design was functionally verified and achieved timing closure at $3\,\mathrm{GHz}$ under typical PVT conditions.

\begin{table}[t]
\centering
\setlength\tabcolsep{6pt}
\caption{Design Area Overhead of the Interconnect routing and PUF in \textsc{InterPUF}.}
\label{tab:interpuf_design_area}
\begin{tabular}{lccc}
\toprule
\textbf{Design} &
\begin{tabular}[c]{@{}c@{}}\textbf{Design Area}\\(\(\mu m^2\))\end{tabular} &
\begin{tabular}[c]{@{}c@{}}\textbf{Interconnect Route Area}\\(\(\mu m^2\))\end{tabular} &
\begin{tabular}[c]{@{}c@{}}\textbf{Overhead}\\(\%)\end{tabular} \\
\midrule
CVA6~\cite{cva6}        &   345{,}755  & 812.0  & 0.23 \\
NVDLA~\cite{nvdla_hw}   &   541{,}552  & 835.5  & 0.15 \\
RISC\text-\!V~\cite{riscv_soc}  & 1{,}309{,}680  & 790.1  & 0.06 \\
Ariane~\cite{ariane}    & 1{,}431{,}536  & 868.3  & 0.06 \\
OR1200~\cite{ARM}       & 1{,}488{,}384  & 805.2  & 0.05 \\
\bottomrule
\end{tabular}
\end{table}

\begin{table}[t]
\centering
\setlength\tabcolsep{6pt}
\caption{Design Area Overhead of Chiplet Encryption (per–chiplet, with proposed changes).}
\label{tab:chiplet_encryption_area}
\begin{tabular}{lccc}
\toprule
\textbf{Design} &
\begin{tabular}[c]{@{}c@{}}\textbf{Design Area}\\(\(\mu m^2\))\end{tabular} &
\begin{tabular}[c]{@{}c@{}}\textbf{Chiplet Encryption Area}\\(\(\mu m^2\))\end{tabular} &
\begin{tabular}[c]{@{}c@{}}\textbf{Overhead}\\(\%)\end{tabular} \\
\midrule
CVA6~\cite{cva6}        &   345{,}755  & 6{,}988.2  & 2.02 \\
NVDLA~\cite{nvdla_hw}   &   541{,}552  & 7{,}192.1  & 1.33 \\
RISC\text-\!V~\cite{riscv_soc}  & 1{,}309{,}680  & 7{,}063.0  & 0.54 \\
Ariane~\cite{ariane}    & 1{,}431{,}536  & 7{,}000.5  & 0.49 \\
OR1200~\cite{ARM}       & 1{,}488{,}384  & 7{,}108.0  & 0.48 \\
\bottomrule
\end{tabular}
\end{table}

\begin{table}[t]
\centering
\setlength\tabcolsep{3pt}
\caption{Area Comparison with Recent Works.}
\label{tab:area_comparison_recent}
\begin{tabular}{lccc}
\toprule
\textbf{Work} & \textbf{LUT Resource} & \textbf{FF} & \textbf{Area (mm$^2$)} \\
\midrule
SECT-HI~\cite{sect_hi} & -- & -- & $5.11$ \\
PQC-HI (Kyber+Dilithium)~\cite{pqc_hi} & $76{,}999$ & $49{,}993$ & --  \\
PQC-HI (Kyber only)~\cite{pqc_hi} & $1{,}842$ & $1{,}634$ & -- \\
\textbf{\textsc{InterPUF} (1 chiplet)} & \textbf{1{,}915} & \textbf{1{,}160} & \textbf{0.0078} \\
\textbf{\textsc{InterPUF} (32 chiplets)} & \textbf{61{,}280} & \textbf{37{,}120} & \textbf{0.2501} \\
\bottomrule
\end{tabular}
\end{table}

\subsection{Area Efficiency and Scalability}
\label{sec:area_eval}

Area efficiency is one of the most critical enablers for security primitives in heterogeneous multi-chiplet systems. A security mechanism that consumes excessive silicon quickly becomes impractical at scale, where dozens of chiplets may be integrated within a single package. The key advantage of \textsc{InterPUF} lies in its architecture; rather than replicating costly encryption engines within each chiplet, authentication is embedded directly into the interconnect fabric. This design ensures that the majority of the security cost is a one-time expense at the SiP level, while the marginal per-chiplet overhead remains negligible. 

Table~\ref{tab:interpuf_design_area} highlights this property, resulting in great area reduction. The routing fabric and embedded PUF together occupy less than $900~\mu m^2$, which translates to an overhead of only $0.05$--$0.23\%$ depending on the SoC baseline. Importantly, this cost does not grow with the number of chiplets in the system: once the secure interconnect is in place, additional chiplets can be authenticated without increasing the routing area. In other words, the fundamental cost of interconnect-level security is fixed for the package, not per chiplet. 

Moreover, summarized in Table~\ref{tab:chiplet_encryption_area}, each chiplet must integrate an encryption core of roughly $7{,}000~\mu m^2$, leading to overheads in the range of $0.5$--$2.0\%$ per design. While such costs may appear modest in isolation, they scale linearly with the number of chiplets. A 32-chiplet system would pay for 32 redundant encryption engines, wasting silicon that could otherwise be allocated to computation or memory. By eliminating these blocks, \textsc{InterPUF} achieves nearly an order-of-magnitude reduction for large-scale systems.

The scalability advantage is further evident when comparing it with recent secure hardware solutions, as shown in Table~\ref{tab:area_comparison_recent}. 
\uline{For a single chiplet, \textsc{InterPUF} requires only $0.0078~\text{mm}^2$, which is already an order of magnitude smaller than PQC accelerators and several times smaller than enclave-based schemes such as SECT-HI}~\cite{sect_hi}. 
When scaled to $32$ chiplets, the total overhead remains just $0.2501~\text{mm}^2$, which is still dramatically lower than competing solutions. This proves that, unlike most alternatives, \textsc{InterPUF} does not collapse under scale: the security mechanism remains compact even for the most demanding chiplet counts.

Taken together, these results establish a compelling case for \textsc{InterPUF}. It introduces a fixed, SiP-wide interconnect overhead of less than $1\%$, avoids the linear per-chiplet penalties of standalone encryption, and scales gracefully to dozens of chiplets without exceeding the area budgets of modern SoCs. No prior work demonstrates this balance of efficiency and scalability, making \textsc{InterPUF} uniquely suited for secure heterogeneous integration at scale.

\subsection{Power Efficiency and Overhead Analysis}
\label{sec:power_eval}

Power overhead is a critical consideration for security primitives in multi-chiplet systems, since any recurring cost directly impacts runtime energy budgets. Table~\ref{tab:interpuf_power_overhead} shows that the interconnect routing and embedded PUF in \textsc{InterPUF} consume less than 0.011 mW across all evaluated SoCs, corresponding to only 0.005\%–0.072\% overhead relative to baseline power budgets that already span tens to hundreds of milliwatts. This extremely low figure is not incidental, but stems directly from the architectural simplicity of the design. Unlike conventional cryptographic accelerators, \textsc{InterPUF} avoids wide datapaths, deep pipelines, and heavy arithmetic. Instead, it relies on compact combinational logic and a handful of low-overhead sequential registers, leading to inherently low switching activity. As a result, the measured dynamic power falls into the sub-10 µW range, well below typical SoC power variation margins, making the overhead effectively invisible in practice. Moreover, Table~\ref{tab:chiplet_encryption_power_overhead} highlights the comparison with per–chiplet encryption blocks, which consume around 0.24 mW each, or 0.2\%–2.0\% overhead depending on the host design. Still scaling largely, even for typical open-source SoC/chiplets. This can be improved further with the inclusion of clock gating or repurposing the encryption core for functional use for the chiplets.

\begin{table}[t]
\centering
\setlength\tabcolsep{4pt}
\caption{Power Overhead of Interconnect Routing in \textsc{InterPUF}.}
\label{tab:interpuf_power_overhead}
\begin{tabular}{lccc}
\toprule
\textbf{Design} & 
\textbf{\begin{tabular}[c]{@{}c@{}}Baseline Power \\ (mW)\end{tabular}} & 
\textbf{\begin{tabular}[c]{@{}c@{}}Interconnect Power \\ (mW)\end{tabular}} & 
\textbf{\begin{tabular}[c]{@{}c@{}}Overhead \\ (\%)\end{tabular}} \\
\midrule
CVA6~\cite{cva6}   & 12.896  & 0.0093 & 0.072 \\
NVDLA~\cite{nvdla_hw}  & 185.140 & 0.0098 & 0.005 \\
RISC-V~\cite{riscv_soc} & 59.164  & 0.0087 & 0.015 \\
Ariane~\cite{ariane} & 94.157  & 0.0102 & 0.011 \\
OR1200~\cite{ARM} & 106.610 & 0.0091 & 0.009 \\
\bottomrule
\end{tabular}
\end{table}

\begin{table}[t]
\centering
\setlength\tabcolsep{4pt}
\caption{Power Overhead of Chiplet Encryption (per–chiplet, with proposed changes).}
\label{tab:chiplet_encryption_power_overhead}
\begin{tabular}{lccc}
\toprule
\textbf{Design} & 
\textbf{\begin{tabular}[c]{@{}c@{}}Baseline Power \\ (mW)\end{tabular}} & 
\textbf{\begin{tabular}[c]{@{}c@{}}Chiplet Encryption Power \\ (mW)\end{tabular}} & 
\textbf{\begin{tabular}[c]{@{}c@{}}Overhead \\ (\%)\end{tabular}} \\
\midrule
CVA6~\cite{cva6}   & 12.896  & 0.2357 & 1.83 \\
NVDLA~\cite{nvdla_hw}  & 185.140 & 0.2394 & 0.13 \\
RISC-V~\cite{riscv_soc} & 59.164  & 0.2518 & 0.43 \\
Ariane~\cite{ariane} & 94.157  & 0.2463 & 0.26 \\
OR1200~\cite{ARM} & 106.610 & 0.2485 & 0.23 \\
\bottomrule
\end{tabular}
\end{table}

In comparison to other works, \textsc{InterPUF} maintains its negligible footprint regardless of scale, and its compact logic is highly amenable to standard low-power techniques such as clock gating and operand isolation, which could reduce consumption even further. 
\uline{Other recent works, such as SECT-HI and PQC-HI, did not report power results, but given their millimeter-scale area and reliance on complex cryptographic primitives, their power overheads are expected to be significantly larger}~\cite{sect_hi,pqc_hi}. 
Therefore, \textsc{InterPUF} provides a much better security solution with virtually negligible power overhead. Combined with its area efficiency, these results demonstrate that \textsc{InterPUF} not only reduces silicon footprint but also ensures energy sustainability for secure heterogeneous integration.

\begin{figure}[t]
    \centering
    \includegraphics[width=\linewidth]{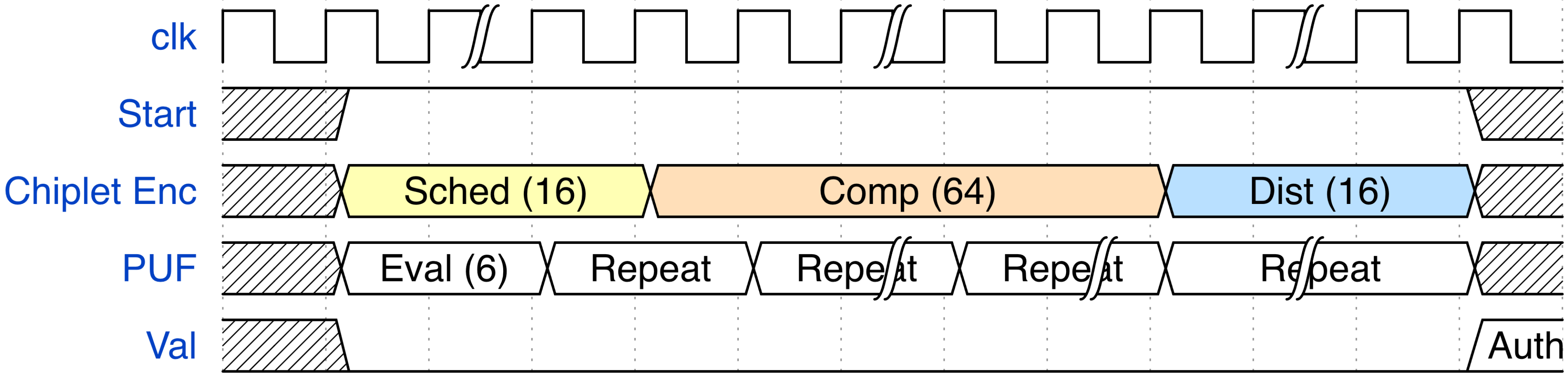}
    \caption{Waveform diagram of authentication stages in \textsc{InterPUF}. The interconnect PUF completes in $6$ cycles and can be repeated multiple times within a single SHA-256 window ($96$ cycles). This ensures that interconnect validation is established as the root of trust before authentication.}
    \label{fig:timing_waveform}
\end{figure}

\subsection{Authentication Latency Breakdown}
\label{sec:latency_eval}

Authentication in \textsc{InterPUF} proceeds in two tightly coupled stages: first, the interconnect fabric is authenticated using the embedded PUF, and then chiplet-level claims are validated using a \textsc{sha256} pipeline. This staged approach ensures that the interconnect, the communication backbone of the SiP, is established as the root of trust before chiplets are admitted into the system.

The interconnect authentication is extremely low-overhead, requiring only $6$ cycles in total: one cycle for challenge scheduling followed by five cycles of PUF evaluation. At our $3\,\mathrm{GHz}$ operating frequency, this corresponds to roughly $2\,\mathrm{ns}$. Because of its small footprint, the interconnect can be re-evaluated many times with virtually no penalty. For instance, during the $96$ cycles required to compute a single \textsc{sha256} digest, the interconnect can be authenticated up to $96/6 \approx 16$ times. This allows repeated testing, majority voting for error suppression, and rotation of path configurations to increase modeling resistance, so the routing fabric is validated with high confidence before chiplet-level authentication begins.

Chiplet authentication is performed by transmitting hashed signatures that are validated through the \textsc{sha256} pipeline. This hashing stage requires $96$ cycles in total, with $16$ cycles for message scheduling and $64$ cycles for compression rounds, followed by minor control overhead, resulting in less than $32\,\mathrm{ns}$ at $3\,\mathrm{GHz}$. This process runs in parallel with ongoing interconnect validation, so by the time the chiplet digest is ready, the fabric has already been re-checked several times.

\rev{To bind each acceptance to the current PUF-derived session, we add a constant-round two-party computation (\textsc{2pc}) step (Yao garbled circuits with Oblivious Transfer). The online cost is dominated by sending one garbled circuit for a fixed function $f_i$ (recompute \textsc{sha256} over the chiplet’s identity material and $R^\star$, check equality with the stored commitment, then derive a session token) plus one batch of \textsc{ot}-extension messages, and finally revealing the outputs. With half-gates, the garbled size for this circuit is on the order of $0.6$--$1.0$\,MB at $128$-bit security; on an on-package link of $32$\,GB/s this serializes in $\approx 20$--$30\,\mu$s, and on a $128$\,GB/s link in $\approx 5$--$8\,\mu$s. The round complexity is constant (two to three exchanges), so latency is bandwidth-bound rather than depth-bound. Thus the end-to-end per-chiplet path is
\[
t_{\text{PUF}} \approx 2\,\text{ns}, \quad
t_{\text{\textsc{sha256}}} < 32\,\text{ns}, \quad
t_{\text{2pc}} \approx 5\text{--}30\,\mu\text{s},
\]
with the nanosecond-scale PUF and hashing stages fully overlapped by the microsecond-scale \textsc{2pc}. Multiple chiplets can be processed in parallel lanes while the interconnect continues periodic PUF self-checks.}

The timing relationship between interconnect and chiplet authentication is illustrated in Fig.~\ref{fig:timing_waveform}: the interconnect completes in a handful of cycles and can be repeated without bubbles, while the \textsc{sha256} pipeline handles chiplet signatures in a single digest round, and the \rev{constant-round \textsc{2pc}} adds a bounded, bandwidth-driven tail. This overlapping structure minimizes end-to-end latency and enforces a clear trust hierarchy, where the interconnect is first secured and chiplets are authenticated against it, providing both speed and robustness for scalable multi-chiplet systems.






\subsection{PUF Simulation and Modeling Attack Results}
\label{subsec:puf_eval}

\begin{figure}[t]
    \centering
    \includegraphics[width=\linewidth]{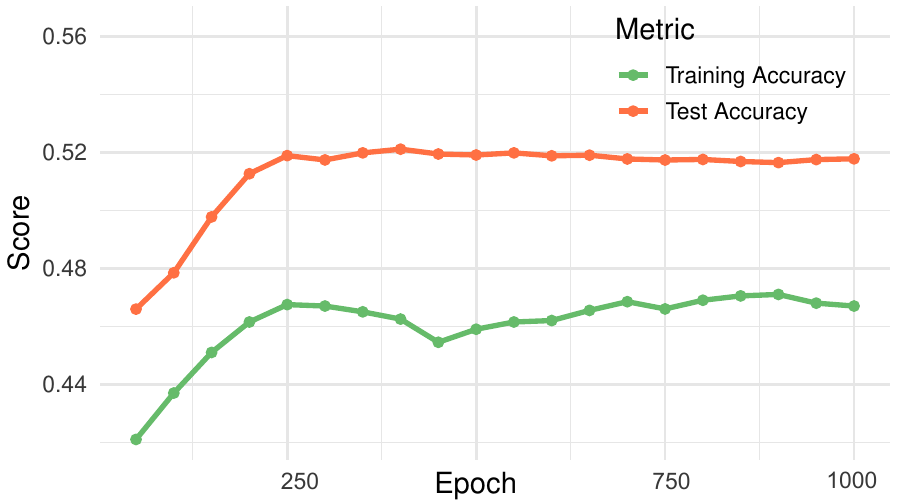}
    \caption{Training curve for the modeling attack on \textsc{InterPUF}, showing training and test accuracy over $1000$ epochs. The training and test accuracy converges near $46\%$ and $52\%$ respectively, confirming resilience against learning models.}
    \label{fig:interpuf_training_curve}
\end{figure}

To assess the robustness of the proposed \textsc{InterPUF}, we conducted detailed simulations using the \textsc{pypuf} framework (Tab.~\ref{tab:pypuf_results}). Five synthetic chip instances were generated under noise and process variation, with challenges transformed through router-based permutations and sparse flips. The evaluation confirmed strong statistical quality: average uniformity was $0.4986 \pm 0.0028$, close to the ideal $0.5$, and bias remained negligible at $0.0049$. Reliability was high at $98.2\%$, and intra-chip Hamming distance across repeated queries was tightly bounded at $0.0189$, confirming stability. Uniqueness, measured as mean inter-chip Hamming distance, was $0.4648$, close to the ideal $0.5$, while bit-aliasing stayed balanced at $0.4986$. Bit-flip sensitivity averaged $0.5140$, consistent with expected avalanche properties. To further test resilience, we applied a machine learning attack using logistic regression trained on $8000$ CRPs and tested on $3000$ unseen CRPs. The model achieved only $46.7\%$ accuracy and an AUC of $0.4899$, which is no better than random guessing (Fig.~\ref{fig:interpuf_training_curve}). This confirms that the transformation and hashing in \textsc{InterPUF} significantly raise the difficulty of predictive modeling, preserving the trust-agnostic guarantees.

\begin{table}[t]
\centering
\caption{Simulation metrics and modeling attack performance (4 chiplets, \textsc{pypuf}).}
\label{tab:pypuf_results}
\setlength{\tabcolsep}{6pt}
\begin{tabular}{lcc}
\toprule
\textbf{Metric} & \textbf{Mean Value} & \textbf{Std. Dev.} \\
\midrule
Uniformity & $0.4986$ & $0.0028$ \\
Bias & $0.0049$ & $0.0037$ \\
Uniqueness (HD) & $0.4648$ & $0.0734$ \\
Reliability & $0.9816$ & -- \\
Intra-chip HD & $0.0189$ & $0.0008$ \\
Bit-aliasing & $0.4986$ & -- \\
Bit-flip sensitivity & $0.5140$ & -- \\
\midrule
Modeling Attack (Accuracy) & $0.4675$ & -- \\
Modeling Attack (AUC) & $0.4899$ & -- \\
\bottomrule
\end{tabular}
\end{table}

\section{Security Analysis}
\label{sec:security}

Security in heterogeneous multi-chiplet systems must be evaluated not only in terms of cryptographic strength, but also with respect to structural resilience against a wide spectrum of practical attacks. Because the interconnect serves as the communication backbone and chiplets themselves may originate from diverse vendors, an effective solution must withstand modeling attempts, replay and downgrade strategies, counterfeit component insertion, package-level tampering, denial-of-service, and side-channel exploitation. This section analyzes how the proposed architecture addresses these threats and demonstrates that robust authentication can be achieved with minimal overhead, while preserving scalability across many-chiplet systems.

\subsection{Modeling and Reliability Attacks}
\label{modeling}

Machine learning attacks on delay-based PUFs typically depend on access to large, stable sets of challenge–response pairs. In this design, two mechanisms frustrate such efforts. First, responses are always masked before being exposed, so attackers never see raw bit flips or stability patterns; second, effective challenges are permuted and masked differently in every session, preventing cross-session dataset aggregation. In addition, repeated evaluation with majority voting eliminates exploitable bias, while attempt gating limits the number of challenges an adversary can collect in practice. Together, these defenses sharply reduce the feasibility of constructing accurate predictive models, as seen in Fig.~\ref{fig:interpuf_training_curve}.

\subsection{Replay, Challenge Modification}
\label{replay}

\begin{figure}[t]
    \centering
    \includegraphics[width=\linewidth]{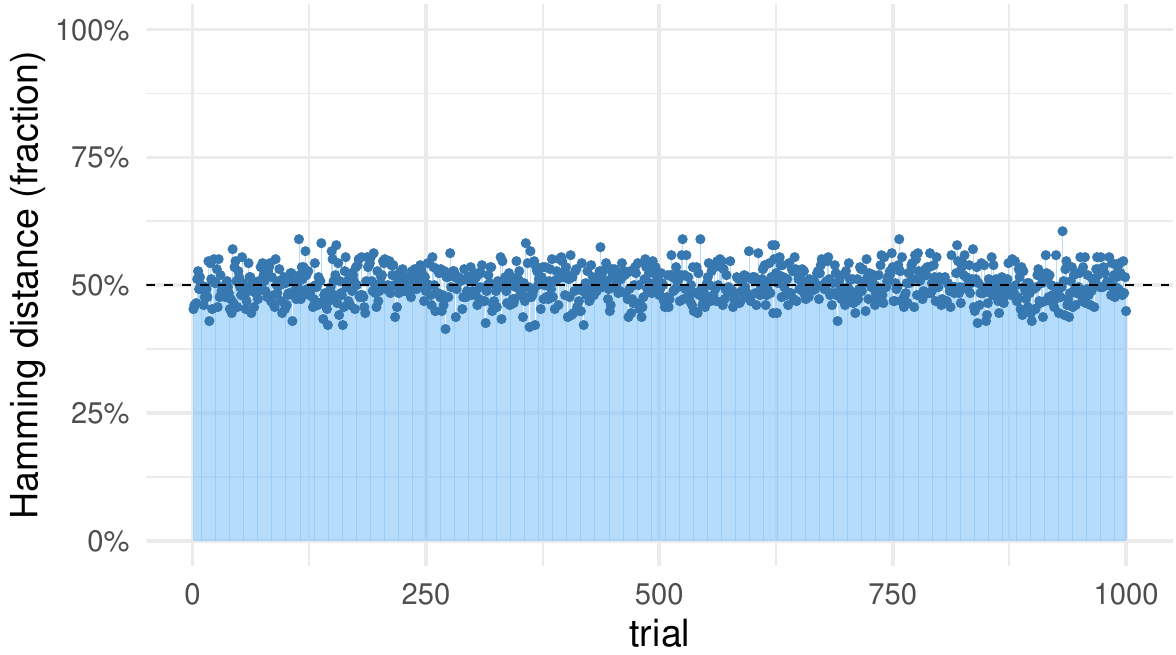}
    \caption{Hamming distance distribution across 1000 trials. The plot demonstrates the stability of tokens, where the distance hovers around 50\%, consistent with strong uniqueness.}
    \label{fig:chiplet_sha_hd}
\end{figure}

Authentication transcripts are bound to per-boot salts and session nonces, ensuring that the same top-level challenge never maps to the same internal state across power cycles. This property renders replay attacks ineffective and previously recorded transcripts are useless in future sessions. Any absence of expected responses or masking signals triggers rejection before chiplets are admitted. Modification attacks also render pointless due to SHA being extremely sensitive (Fig~\ref{fig:chiplet_sha_hd}).

\subsection{Malicious or Counterfeit Chiplets}
\label{counterfeit}

A rogue chiplet may attempt to spoof a valid identity or manipulate interconnect traffic to bias authentication. This system prevents such attacks by authenticating the interconnect before chiplets are considered. Because the backbone is secured first, any attempt to insert a counterfeit component occurs only after the interconnect has been verified multiple times. Once authenticated, genuine chiplets transmit hashed signatures that are validated against reference digests. Since replayed or fabricated signatures cannot pass verification without the correct secret context, malicious chiplets are effectively excluded.

\subsection{Interposer and Package-Level Attacks}
\label{package}

At the package level, an adversary could attempt to emulate valid delay paths by inserting programmable delays or rerouting signals through an interposer. However, the dynamic hashing of challenges and responses makes this infeasible: an attacker would need to replicate the per-boot remapping and masking keyed to device-unique values in real time, across many parallel evaluations. Repeated proofs within a single authentication session significantly increase path coverage, exposing even subtle deviations introduced by interposer-level manipulation.

\subsection{Denial-of-Service and Throughput Abuse}
\label{dos}

Attackers may attempt to flood the system with challenges or intentionally stall transactions. Authentication logic enforces bounded attempts and cooldowns, preventing resource exhaustion. Because each interconnect proof completes in only a handful of cycles, the system can sustain frequent re-checks while maintaining throughput. Moreover, chiplet-level authentication runs concurrently with interconnect checks, ensuring that progress is not blocked even under partial denial-of-service attempts.

\subsection{Side-Channels and Fault Injection}
\label{sidechannel}
Since raw responses are never exposed, the information leaked through power or timing side-channels is already obfuscated and session-dependent. The logic itself is narrow, with constant-latency control paths and very low switching activity, reducing exploitable leakage compared to large cryptographic blocks. Established countermeasures such as clock gating, operand isolation, and randomized re-check cadence can further strengthen resilience if required. Fault injection or environmental drift is mitigated by majority voting across repeated evaluations, allowing the system to reject inconsistent results and remain robust under variation.

\subsection{Removal Attacks}
\label{removal}

In chiplet-based integration, an adversarial foundry or vendor may try to strip away authentication circuitry, bypass it, or insert malicious modifications. In \textsc{InterPUF}, resilience is ensured by embedding authentication directly into the interconnect fabric rather than treating it as a detachable add-on. Because the security primitives are interwoven with routing and switching logic, there is no clean boundary to isolate and remove. Responses are further obfuscated and blended with normal interconnect activity, making them indistinguishable from functional signals. Any modification disrupts communication consistency and is quickly exposed during repeated challenge–response checks, making removal attacks significantly harder.

\section{Discussion and Limitations}
\label{sec:discussion}

\textsc{InterPUF} demonstrates several strengths.~\circled{\small 1}~It achieves authentication with negligible overhead in both area and power,~\circled{\small 2}~scales efficiently with growing chiplet counts, and~\circled{\small 3}~enforces a strict minimal-trust model by combining interconnect-level PUF proofs with chiplet-level hashed signatures. These properties make it low-overhead, practical, and well-suited for heterogeneous integration compared to prior centralized or heavyweight cryptographic approaches. Nonetheless, limitations remain. Our evaluation is confined to RTL-level prototypes and simulations, so real silicon validation under broader PVT variations, aging, and environmental stress is still required to confirm robustness over time. \rev{We have not yet built or evaluated an \textsc{fpga} or silicon prototype; both are planned as next steps to validate functionality, quantify stability across PVT/aging, and measure end-to-end throughput and latency on hardware.} \rev{A tape-out with controlled process corners and an \textsc{fpga}-based emulation of the authentication controller are on our roadmap to complement the current synthesis and simulation-based results.}

\section{Conclusion}
\label{sec:conclusion}

\textsc{InterPUF} introduces a low-overhead, interposer-resident authentication framework that combines delay-based PUFs with multi-party computation to enforce a strict minimal-trust model in heterogeneous chiplet systems. By embedding the root of trust in the interconnect and layering it with chiplet-level cryptographic signatures, the scheme achieves strong resilience against modeling, replay, and counterfeit attacks with negligible area and power overhead. Our RTL implementation and simulations confirm that the design scales efficiently with chiplet count, providing a practical path toward secure, reconfigurable system-in-package integration.

\bibliographystyle{IEEEtran}
\bibliography{sample-base}

\end{document}